\documentstyle[amscd,amssymb,verbatim,12pt]{amsart}
\def\dspace{\baselineskip=0.3 in}
\begin{document}
\dspace
\title[Non-linear equation of state,cosmic acceleration and ....]{Non-linear equation of state,cosmic acceleration and deceleration during phantom-dominance }

\author[S.K.Srivastava and J. Dutta]%
        {    }

\maketitle

\centerline{\bf S.K.Srivastava$^1$ and J. Dutta$^{1,2}$ }

\centerline{$^1$ Department of Mathematics, North Eastern Hill University,}

\centerline{ NEHU Campus,Shillong - 793022 ( INDIA ) }

\centerline{and }

\centerline{$^2$ Department of Mathematics, St. Edmund's College}

\centerline{ Shillong-793003 ( INDIA ) .}

\vspace{0.5cm}

\centerline{\bf Abstract}

\smallskip

Here, RS-II model of brane-gravity is considered for phantom universe using
non-linear equation of state. Phantom fluid is known to violate the weak
energy condition. In this paper, it is found that this characteristic of
phantom energy is affected drastically by the negative brane-tension $\lambda$ of
the RS-II model. It is interesting to see that upto a certain value of energy
density $\rho$ satisfying $\rho/\lambda < 1$, weak energy condition is
violated and universe super-accelerates. But as $\rho$ increases more, only
strong energy condition is violated and universe accelerates.

----------------------------------------------

Electronic addresses-

S.K.Srivastava : srivastava@@nehu.ac.in ; sushilsrivastava52@@gmail.com

J.Dutta : jibi\_ dutta@@yahoo.co.in

\newpage

\noindent  When $1 < \rho/\lambda < 2$, even strong energy condition is not
violated and universe decelerates. Expansion of the universe stops , when $\rho= 2
\lambda.$  This is contrary to earlier results of phantom universe
exhibiting acceleration only.


\vspace{1cm}

\centerline{\bf 1.Introduction}

\smallskip

Astrophysical observations,  in the recent past, have conclusive evidence
in favor of 
late cosmic acceleration\cite{sp, sp03, ag}. This revolutionary observation 
challenged cosmologists to develop an appropriate cosmological model 
explaining acceleration in the late universe. Also, observations support
homogeneous and flat model of the universe \cite{ad}. In such a
model, Friedmann equations show that cosmic dynamics can
exhibit acceleration only when $\rho + 3 p < 0$ with
$\rho$ being the energy density and $p$ being the pressure density
\cite{ejc}. It shows violation of the cosmic strong energy condition (SEC) and
 indicates dominance of exotic fluid in the late universe. The condition $\rho
 + 3 p < 0$ implies  $- 1/3
> w > - 1$ with EOS parameter $w = p/\rho.$ In 2002-2003, Caldwell found the
case $w < - 1$ 
better fit for the observed astrophysical data and  advocated for this case,
which violates the weak energy condition (WEC) too \cite{rr, rr03}. This fluid,
 is known as phantom. The phantom model
 explains the present and future acceleration of the universe, but it is
 plagued with the problem of big-rip singularity ( singularity in finite
 future time when energy density, pressure and the scale factor
 diverge). Thus, phantom  was another exotic matter suggested by Caldwell.
Different sources of exotic matter
violating SEC \cite{jm, ca, ttm, as, mrg, psm, sk04, ms, gcs} and WEC were proposed in the recent
past \cite{rr, rr03, sk,  rj,ob,mc, sc, sc03, sm,add, mrw, sno, sno03, abd,
  om,ka, mb}. A comprehensive review of these contributions is
available in \cite{ejc}. Later on, it was proposed that curvature could be a
possible source of dark energy. These models are known as $f(R)-$ dark energy models \cite[for
review]{sn06}. Recently, $f(R)-$ dark energy models are criticized in
\cite{lds, lds07} on the ground 
that these models  do not produce matter in
the late universe needed for formation of large scale structure in the
universe. In another review, Nojiri and Odintsov have discussed dark matter in
the late universe refuting criticism against their work \cite{sn08}. Apart from these, curvature-induced dark energy models, different
from $f(R)-$ dark energy models, were proposed in \cite{sks, sks06, sks07}.

In the race to investigate a viable cosmological model, satisfying
observational constraints and explaining present cosmic acceleration, brane-gravity was
also drawn into service and brane-cosmology was developed. A review on
brane-gravity and its various applications with special attention to cosmology
is available in\cite{var, rm,   pbx, cs}. 
After development of M-theory bringing different string-theories under one
umbrella, this theory stemmed from low energy string theory when Ho\v{r}ova and Witten proposed that 11-dimensional supergravity,
being a supermembrane theory, could be obtained as low-energy limit of 11-dimensional
M-theory. They discussed that it could be done  on a particular orbifold $R^{10}
\otimes S^1/Z_2$ with $R^{10}$ being the 10-dimensional space-time and
$S^1/Z_2$ being the 1-dimensional space having  $ x^{11} \leftrightarrows -
x^{11}$ symmetry \cite{ho}. According to this solution, when the six extra-dimensions on $(1 + 9)$-branes are
compactified on very small scale close to the fundamental scale, their effect
is realized on $(1 + 3)$-dimensional brane located at the ends of $ S^1/Z_2$.  Thus,  Ho\v{r}ova - Witten solution provided an effective 5-dimensional model
where extra-dimension can be large relative to the fundamental scale in
contrast to Kaluza-Klein theory, where extra-dimension is very small
\cite{app, skp}. This
solution was used by L.Randall and R.Sundrum in their seminal paper to solve
the ``hierarchy problem'' by a {\em warped} or curved dimension showing that
fundamental scale could be brought down from the Planck scale to 100 GeV.  Thus,
Randall-Sundrum approach brought the theory to scales below 100 GeV being
the electroweak scale( so far results could be verified experimentally upto
this scale only). In this model, extra-dimension is
large having $(1 + 3)$-branes at its ends. These branes are $Z_2$-symmetric (have
mirror symmetry) and have tension to counter the negative cosmological
constant in the ``bulk'', which is ${\rm AdS}_5$. The model, having {\em two}
$(1 + 3)$-branes at the ends of the orbifold $S^1/Z_2$, is known as RS-I model \cite{rs1}.

In another paper, in the same year, these authors proposed another brane-model
as an alternative to compactification . In this model, there is only {\em one}
$(1 + 3)$-brane at one end of the extra-dimension and the other end tends to
infinity. This model is known as RS-II model \cite{rs2}. Thus, Ho\v{r}ova -
Witten solution and RS- theory yielded
brane-gravity originating from low-energy string theory, which  explained
weakness of gravity also in the observable
universe. In case, the extra-dimension is time-dependent, brane-gravity induced
Friedmann equation (giving dynamics of the universe) contains a correction
term $- {4\pi G
  \rho^2}/{3\lambda}$  with $\lambda$ being the
brane-tension \cite{var, rm,   pbx, cs} . In RS-I
model, $\lambda$ is positive, whereas  $\lambda$ is negative in RS-II
model. 

So, apart from general relativity (GR)-based models and $f(R)-$ models,
brane-gravity (BG)-based cosmological models were also tried upon to explain
acceleration in the  late universe. In particular, RS-II model got much attention due to its simple
and rich conceptual base \cite{pb00, pbd00, cg, cgs, mwb, cll, cal, apt, nopl, sksgrg}. In \cite{vs3}, it is found that RS-II
model of brane-gravity yields a phantom model giving transient acceleration
(where acceleration stops after sometime in future) and avoiding big-rip
singularity. Later on, avoidance of big-rip singularity  was shown in GR-based model too , if the dominating fluid bahaves as barotropic phantom fluid
and generalized Chaplygin gas simultaneously \cite{sk}. 
 
 The present density of dark
energy is found to be $0.73 
\rho^0_{\rm cr}$ with $ \rho^0_{\rm cr} = 2.5
\times 10^{-47} {\rm GeV}^4$ (present critical energy density )\cite{sp,
  sp03}. If late universe is dominated 
by phantom , present phantom energy density is $0.73
\rho^0_{\rm cr}$. As discussed above, Friedmann equation, obtained from RS-II
model of brane-gravity, contains the energy term as $(8\pi G \rho/3)[1 -
(\rho/2\lambda)]$, where $\lambda = 48
\pi G/  k_5^4 = 48
\pi/ M_P^2 k_5^4 $ with $k_5^2 = 8 \pi G_5 = 8 \pi G l =  8 \pi l/M_P^2,$ Newtonian
gravitational constant $G = M_P^{-2}$ in natural units given below ($M_P$ being
the Planck mass) and $l$ being 
length of the extra-dimsion of the 5-dimensional bulk\cite{rm,   pbx, cs}.  As an example, if
we set $k_5^2 = 1 {\rm GeV}^{-3}$  as taken in
ref.\cite{pbx}, $\lambda = 48
\pi/ M_P^2 = 6.03 \times 10^{10} \rho^0_{\rm cr} $ and $\rho/2\lambda \sim
10^{-10}$ 
in the present universe . This example shows that
the brane- correction term, in the Friedmann equation, may not be effective in
the present universe unless length of the
extra-dimension is sufficiently small. But, energy density of the phantom fluid will increase with
expansion of the universe due to EOS parameter $w < -1$ in this case. So, even if the correction term is not
effective in the present universe, it will be effective in future phantom
universe. 

In what follows, {\em three} situations are obtained. In
the present universe, brane-corrections are not effective (as obtained above). This situation will
continue until $\rho$ will grow sufficiently. During this period WEC is violated
and phantom universe will super-accelerate. As far as $\rho << 2\lambda$,
universe will super-accelerate in future and $\rho$ will grow with $a(t)$. It
is reasonable to believe  $\rho/\lambda \gtrsim 1$ in future due to
rapid increase in $a(t)$ caused by super-acceleration. Increase in $\rho$ will
still continue with growing 
$a(t)$. On further increase in $\rho$,
brane-corrections will be effective and only SEC
will be violated upto a certain value of phantom energy density. As a consequence, acceleration of the phantom universe will
become comparatively slow. It means that, in this situation, phantom universe
will accelerate, but it will not super-accelerate. It is because,  ${\ddot
  a}/a >  8\pi G \rho/3$ when WEC is violated and $0 <{\ddot a}/a < 8\pi G
\rho/3 ,$  when only SEC is violated. This is the intermediate
state. When $\rho$ will increase more, none of SEC and WEC will be violated
due to strong effect of brane-corrections and phantom universe will
decelerate.  Acceleration and super-acceleration manifest anti-gravity effect
of dark energy. So, it is found that that brane-corrections, in RS-II model,
counter anti-gravity effect of phantom dark energy. In the case of
quintessence, $\rho$ decreases with expansion of the universe due to $w > -
1$, so brane-corrections can not be effective in RS-II model-based present and
future quintessence universe.

As discussed above, it is found that $w = -1$ divides the cases violating SEC
and WEC. It is known as phantom divide. It means that ideal EOS $p = - \rho$
needs a correction term  being linear or non-linear function of $\rho$ causing
deviation from the ideal situation as $p = - \rho \pm f(\rho)$. Here, only
phantom fluid is considered, so we take the negative sign  yielding $w <
-1$. Moreover, $f(\rho)$ in the proposed non-linear EOS implies dependence of
$w$ on $\rho$. In some earlier investigations \cite{not, st, no}, these types of EOS were used
 considering time-dependent viscosity with
correction terms dependent on $\rho$ and $ H = {\dot a}/a ({\dot
  a} = da/dt)$. 

In a recent paper \cite{sk07},  EOS $p = - |w| \rho$ for phantom
fluid with constant $w < -1$ has been considered in RS-II
model based Friedmann equation and it is found that brane-gravity corrections
suppress the phantom characteristic to violate WEC and SEC, when $\rho$
increases sufficiently with expansion of the universe.  As
a consequence, this model expands with acceleration upto some finite time
explaining present cosmic acceleration , but it
decelerates later on. In paper \cite{sk07}, $f(\rho)$ is a linear
 function of $\rho$. So, it is natural to study effect of brane-corrections
 taking non-liear $f(\rho)$ too. Aim
of the present paper is to extend the work of \cite{sk07} taking EOS $p = -
\rho - f(\rho)$ with $f(\rho)$  being non-linear functional of $\rho$.

The paper is organized as follows. In section 2, effective EOS is obtained
with brane-gravity corrections. Section 3 contains discussion on acceleration
and deceleration of the model. Section 4 summarizes the work.

 Here, natural units  $ {\hbar}=c=1$ are used, where ${\hbar}$ and $c$ have their standard meaning.

\bigskip

\centerline{\bf 2. Effective equation of state}

Observations support homogeneous and isotropic model of the late universe,
given by the line-element \cite{ad}
 $$ ds^2 = dt^2-a^{2}(t)[dx^2 +dy^2 + dz^2]   \eqno(2.1)$$ 
where a(t) is the scale factor.

In this space-time, RS-II model based  Friedmann equation is obtained as
\cite{rm, nopl, sksgrg} 
$$ H^2 = {\Big( \frac{\dot a}{a}\Big)}^2 =\frac{8\pi G}{3}\rho\Big
[1-\frac{\rho}{2\lambda}\Big]    \eqno(2.2)$$ 
with $G, \rho $ and $\lambda$ defined above. 

In the brane-cosmology also, conservation equation is given as \cite{rm}
$$ {\dot \rho} + 3H(\rho + p) = 0 .    \eqno(2.3)$$

Connecting (2.2) and (2.3), it is obtained that \cite{nopl, sksgrg}
  $$ \frac{\ddot a}{a}=-4\pi G(\rho
  +p)\Big[1-\frac{\rho}{\lambda}\Big]+\frac{8\pi
    G}{3}\rho\Big[1-\frac{\rho}{2\lambda}\Big]  \eqno(2.4)$$ 
Here, the non-linear equation of state for phantom fluid is taken as 
$$ p =-\rho - f(\rho) \eqno(2.5)$$

As $\rho + p < 0,$ (2.3) yields ${\dot \rho} > 0$. It shows that phantom
energy density will increase in future with growing $a(t)$. Moreover, (2.2)
shows that $a(t)$ will be maximum when  $\rho = 2 \lambda$.

In GR-based theory, Friedmann equation is obtained as
$$\frac{\ddot a}{a} = - \frac{4\pi G}{3} [\rho + 3 P] . \eqno(2.6)$$

Comparing (2.4) and (2.6), {\em effective EOS} with brane gravity
corrections is obtained as
$$ P = - \rho - f \Big[1 - \frac{\rho}{\lambda}\Big] +
\frac{\rho^2}{3\lambda} \eqno(2.7)$$
using eq.(2.5).

(2.7) yields 
$$ \rho + P = - f\Big[1 - \frac{\rho}{\lambda}\Big] +
\frac{\rho^2}{3\lambda} \eqno(2.8)$$
and
$$ \rho + 3 P = - 2\rho - 3{\tilde f} \Big[1 - \frac{\rho}{\lambda}\Big] +
\frac{\rho^2}{\lambda}.  \eqno(2.9)$$

It is interesting to see from (2.8) and (2.9)  that
 $$\rho + P = -  f \Big[1 + \frac{\rho}{\lambda}\Big] -
\frac{\rho^2}{3\lambda}< 0 $$ 
showing violation of WEC
and
$$ \rho + 3 P = - 2\rho - 3 f \Big[1 + \frac{\rho}{\lambda}\Big] -
\frac{\rho^2}{\lambda} < 0  $$
showing violation of SEC, if $\lambda > 0$, which is the case of RS-I
model. But, in the RS-II model being addressed here, we find certain
situations when these cosmic conditions are not violated due to effect of
brane-gravity corrections. 

Here, the case of RS-II model is analyzed  taking following three cases
yielding different non-linear EOS (2.5) for the phantom fluid  
\begin{eqnarray*}
 (I) f(\rho)&=& A{\rho}^\alpha \\
(II) f(\rho)&=&\frac{{A\rho^\alpha}}{{\sqrt{1-\frac{\rho}{2\lambda}}}} \\
 (III)
 f(\rho)&=&\frac{A\rho^{\frac{1}{2}}ln(\rho/\rho_0)}{\sqrt{1-\frac{\rho}{2\lambda}}}    
\end{eqnarray*}
with $\alpha$ being a real number and $A$ being a coupling constant having
dimension $(mass)^{4-4\alpha}$ in cases(I) and (II). Moreover, $A$ has  dimension $(mass)^2$  in case(III).

\[
  \underline{{\rm Case I}:  f(\rho) = A{\rho}^{\alpha}}
\] 

In this case, (2.8) implies that 
$$ P = - \rho - A {\rho}^{\alpha} \Big[1 - \frac{\rho}{\lambda}\Big] +
\frac{\rho^2}{3\lambda} .\eqno(2.10)$$
(2.10) yields effective pressure $P < 0$ for
$$ \rho < 3 \lambda \Big[ 1 + A {\rho}^{({\alpha} - 1)}\Big\{1 -
\frac{\rho}{\lambda}\Big\}\Big]. \eqno(2.11)$$

Further, it is found that
$$\rho + P = - A {\rho}^{\alpha} \Big[1 - \frac{\rho}{\lambda}\Big] +
\frac{\rho^2}{3\lambda} < 0$$
till
$$ \rho_0 <\rho < 3\lambda A {\rho}^{({\alpha} - 1)} \Big[1 -
\frac{\rho}{\lambda}\Big] \eqno(2.12)$$
with $ \rho_0$ being the present energy density.
This result shows that WEC will be violated till $\rho$ will satisfy the inequality
(2.12). It will not be violated when
$$ \rho > 3\lambda A {\rho}^{({\alpha} - 1)} \Big[1 -
\frac{\rho}{\lambda}\Big]. \eqno(2.13)$$

It means that phantom fluid will behave effectively as
phantom dark energy till $\rho$ will
obey the inequality (2.12). It will  not behave
effectively as phantom when $\rho$ will increase more and will obey the inequality (2.13). 

Moreover,(2.9) shows that SEC will be  violated till
$$  \rho < \lambda \Big[2 +  3 A \rho^{(\alpha - 1)}
  \Big\{1-\frac{\rho}{\lambda}\Big\}  \Big]  .  \eqno(2.14)$$

It shows that when $\rho$ will increase such that
$$ 3\lambda A {\rho}^{({\alpha} - 1)} \Big[1 -
\frac{\rho}{\lambda}\Big] < \rho <
 \lambda \Big[2 +  3 A \rho^{(\alpha - 1)}
\Big\{1-\frac{\rho}{\lambda}\Big\}  \Big],   \eqno(2.15)$$ 
only SEC will be violated. It shows that when $\rho$ will satisfy the
inequality (2.15), phantom 
  characteristic to violate WEC will be suppressed by brane-gravity effects for
  negative brane tension and phantom fluid will behave effectively as
  quintessence. These 
  results yield {\em effective phantom divide} at 
$$\rho = \rho_{\rm phd} = 3\lambda A {\rho}_{\rm phd}^{({\alpha} - 1)} \Big[1 -
\frac{\rho_{\rm phd}}{\lambda}\Big]. \eqno(2.16)$$

It is interesting to see that even SEC will not be violated when
$$ \lambda \Big[2 +  3 A \rho^{(\alpha - 1)} \Big\{1-\frac{\rho}{\lambda}\Big\}  \Big] < \rho < 2 \lambda. \eqno(2.17)$$
implying that, during the range (2.17), dark energy characteristic  to violate
  SEC and WEC will be suppressed completely by brane-corrections in RS-II model.

Thus, the above analysis shows that universe will accelerate till $\rho$ will satisfy
the inequality (2.14) and it will decelerate during the range of $\rho$ given by (2.17). 

\smallskip

\[
  \underline{{\rm Case II}: f(\rho) = {{A\rho^{\alpha}}}/{{\sqrt{1-\frac{\rho}{2\lambda}}}}}
\] 

In this case, (2.8) implies that 
$$ P = - \rho - \frac{{A\rho^\alpha}}{{\sqrt{1-\frac{\rho}{2\lambda}}}} \Big[1 - \frac{\rho}{\lambda}\Big] +
\frac{\rho^2}{3\lambda} .\eqno(2.18)$$
This equation yields effective pressure $P < 0$ for
$$ \rho < 3 \lambda \Big[ 1 +  \frac{{A\rho^{(\alpha
      -1)}}}{{\sqrt{1-\frac{\rho}{2\lambda}}}}\Big\{1 -
\frac{\rho}{\lambda}\Big\}\Big] . \eqno(2.19)$$

Further, it is found that
$$\rho + P = - \frac{{A\rho^\alpha}}{{\sqrt{1-\frac{\rho}{2\lambda}}}} \Big[1 - \frac{\rho}{\lambda}\Big] +
\frac{\rho^2}{3\lambda} < 0 $$
till
$$\rho_0 < \rho < 3\lambda \frac{{A\rho^{\alpha}}}{{\sqrt{1-\frac{\rho}{2\lambda}}}}
\Big[1 - \frac{\rho}{\lambda}\Big]. \eqno(2.20)$$
This result shows that WEC is violated till $\rho$ will satisfy the inequality
(2.20). It will not be violated when
$$ \rho > 3\lambda A {\rho}^{({\alpha} - 1)} \Big[1 -
\frac{\rho}{\lambda}\Big]. \eqno(2.21)$$

So like the case I, in this case too, we find that phantom fluid will not
behave effectively as phantom when $\rho$ will satisfy the inequality (2.21) due
to brane-corrections.

Moreover,(2.9) shows that SEC will be violated till
$$  \rho < \lambda \Big[2 +  \frac{3 A \rho^{(\alpha- 1)}}
  {\sqrt{1 - {\rho}/{2\lambda}} } \Big\{1-\frac{\rho}{\lambda}\Big\}
  \Big], \eqno(2.22)$$ 
but as $\rho$ will increase with time such that
$$ 3\lambda \frac{{A\rho^{\alpha}}}{{\sqrt{1-\frac{\rho}{2\lambda}}}}
\Big[1 - \frac{\rho}{\lambda}\Big] < \rho <  
\lambda \Big[2 +  \frac{3 A \rho^{(\alpha- 1)}}
  {\sqrt{1- {\rho}/{2\lambda}} } \Big\{1-\frac{\rho}{\lambda}\Big\}
  \Big] ,   \eqno(2.23)$$
only SEC will be violated. It shows that when $\rho$ will satisfy the inequality (2.23), phantom
  characteristic to violate WEC will be suppressed by brane-gravity effects for
  negative brane tension and phantom fluid will behave effectively like quintessence. These
  results suggest {\em effective phantom divide} at 
$$\rho = \rho_{\rm phd} = 3\lambda A {\rho}^{({\alpha} - 1)}_{\rm phd} \Big[1 -
\frac{\rho_{\rm phd}}{\lambda}\Big].   \eqno(2.24)$$

It is interesting to see that even SEC will not be violated for
$$ \lambda \Big[2 +  \frac{3 A \rho^{(\alpha- 1)}}
  {\sqrt{1- {\rho}/{2\lambda}} } \Big\{1-\frac{\rho}{\lambda}\Big\}
  \Big] 
  < \rho < 2 \lambda. \eqno(2.25)$$
implying that, during the range (2.25), dark energy characteristic  to violate
  SEC and WEC will be  suppressed completely by brane-corrections in RS-II model.

Thus the above analysis shows that universe will accelerate till $\rho$ will
satisfy the inequality (2.22) and it will decelerate during the range of
$\rho$ given by (2.25).

\smallskip

\[
  \underline{{\rm Case III} :  f(\rho) =
  {A\rho^{1/2}ln(\rho/\rho_0)}/{\sqrt{1-\frac{\rho}{2\lambda}}}} 
\]

In this case, (2.8) implies that 
$$ P = - \rho - \frac{A\rho^{1/2}ln(\rho/\rho_0)}{\sqrt{1-\frac{\rho}{2\lambda}}} \Big[1 - \frac{\rho}{\lambda}\Big] +
\frac{\rho^2}{3\lambda} .\eqno(2.26)$$
This equation yields effective pressure $P < 0$ for
$$ \rho < 3 \lambda \Big[ 1 +  \frac{A\rho^{1/2}ln(\rho/\rho_0)}{\sqrt{1-\frac{\rho}{2\lambda}}}  \Big\{1 -
\frac{\rho}{\lambda}\Big\}\Big] . \eqno(2.27)$$

Further, it is found that
$$\rho + P = -
\frac{A\rho^{1/2}ln(\rho/\rho_0)}{\sqrt{1-\frac{\rho}{2\lambda}}}  \Big[1 - \frac{\rho}{\lambda}\Big] +
\frac{\rho^2}{3\lambda} < 0$$
till
$$\rho_0 <  \rho < 3\lambda  \rho^{-1/2} \frac{A ln(\rho/\rho_0)}{\sqrt{1-\frac{\rho}{2\lambda}}} \Big[1 - \frac{\rho}{\lambda}\Big]. \eqno(2.28)$$
This result shows that WEC will be violated till $\rho$ wil satisfy the inequality
(2.28). It will not be violated when

$$ \rho > 3\lambda  \frac{A\rho^{-1/2}ln(\rho/\rho_0)}{\sqrt{1-\frac{\rho}{2\lambda}}}
\Big[1 - \frac{\rho}{\lambda}\Big]. \eqno(2.29)$$

So like the case I and II, in this case also, we find that phantom fluid will not
behave effectively as phantom when $\rho$ will satisfy the inequality (2.29) due
to brane-corrections.

Moreover, (2.9) shows that, in this case,  SEC will be violated till
$$  \rho < \lambda \Big[2 + \frac{3 A \rho^{- 1/2}ln(\rho/\rho_0)}
  {\sqrt{1- {\rho}/{2\lambda}} } \Big\{1-\frac{\rho}{\lambda}\Big\} \Big] .  \eqno(2.30)$$

It shows that as $\rho$ will increase with time  such that
$$3\lambda  \frac{A\rho^{-1/2}ln(\rho/\rho_0)}{\sqrt{1-\frac{\rho}{2\lambda}}}
\Big[1 - \frac{\rho}{\lambda}\Big]  < \rho   
< \lambda \Big[2 + \frac{3 A \rho^{- 1/2}ln(\rho/\rho_0)}
  {\sqrt{1- {\rho}/{2\lambda}} } \Big\{1-\frac{\rho}{\lambda}\Big\} \Big] ,
  \eqno(2.31)$$ 
only SEC will be violated. It shows that when $\rho$ will satisfy the
inequality (2.31), phantom 
  charactrisic to violate WEC will be suppressed by brane-gravity effects for
  negative brane tension and phantom fluid will behave effectively as
  quintessence. These 
  results suggest {\em effective phantom divide} at
$$\rho = \rho_{\rm phd}^{3/2} = 3\lambda  \frac{A ln(\rho_{\rm
    phd}/\rho_0)}{\sqrt{1-\frac{\rho_{\rm phd}}{2\lambda}}} \Big[1 -
\frac{\rho_{\rm phd}}{\lambda}\Big]. \eqno(2.32)$$

It is interesting to see that even SEC will not be violated for
$$ \lambda \Big[2 + \frac{3 A \rho^{- 1/2}ln(\rho/\rho_0)}
  {\sqrt{1- {\rho}/{2\lambda}} } \Big\{1-\frac{\rho}{\lambda}\Big\} \Big]  <
  \rho < 2 \lambda. \eqno(2.33)$$ 
implying that, during the range (2.33), dark energy characteristic  to violate
  SEC and WEC will be  suppressed completely by brane-corrections in RS-II
  model.

Thus, the above analysis shows that universe will accelerate till $\rho$
satisfies the inequality (2.38) and it will decelerate during the range of
$\rho$ given by (2.33). 

\smallskip

\centerline{\bf 3. Cosmic expansion with acceleration and deceleration}

\centerline{\bf in   RS-II model}

In the preceding section, we obtained different conditions for changes in the behaviour of
phantom fluid dominating the RS-II model-based universe due to brane-gravity
corrections. In what follows, we derive scale factor $a(t)$ solving Friedmann
equation (2.2) and conservation equation (2.3). It helps to find time period,
during which, WEC and SEC will be violated and time period, during which,
these will not be violated.

\smallskip

\[
  \underline{{\rm Case I}:  f(\rho) = A{\rho}^{\alpha}}
\] 
 
In this case, connecting (2.3) and (2.5) ,we obtain
$$ \dot \rho -3A\frac{\dot a}{a}\rho^\alpha = 0 .  \eqno(3.1)$$ 
It integrates to 
$$ \rho=\Big[\rho_0^{1-\alpha} + 3A(1-\alpha)ln(a/a_0)
\Big]^{\frac{1}{1-\alpha}},    \eqno(3.2) $$ 
where $\rho_0\leqslant\rho\leqslant 2\lambda $.

(2.2),(2.3) and (2.5) yield
$$ \dot \rho - 3A\sqrt{\frac{8\pi G}{3}}\rho^{\alpha
  +\frac{1}{2}}\sqrt{1-\frac{\rho}{2\lambda}}= 0 ,    \eqno(3.3) $$ 
where $\rho_0\leqslant\rho\leqslant 2\lambda .$

Exact solution of this equation is obtained as
 $$ t = \frac{1}{A\sqrt{24\pi G}}\Big[t_0 + 2{(2\lambda)}^{(1/2)-\alpha}\Big\{\sqrt{1-\frac{\rho_0}{2\lambda}} 2F_1\Big(\frac{1}{2},\frac{1}{2}+\alpha,\frac{3}{2},1-\frac{\rho_0}{2\lambda}\Big)$$
$$-\sqrt{1-\frac{\rho}{2\lambda}} 2F_1\Big(\frac{1}{2},\frac{1}{2}+\alpha,\frac{3}{2},1-\frac{\rho}{2\lambda}\Big)\Big\}\Big] , \eqno(3.4) $$ 
  where $ 2F_1(a,b,c,x)$ is the hypergeometric function. Further, using (3.2) in (3.4), we get a relation between time $t$ and the scale factor $a(t)$.

As maximum value of $\rho$ is $2\lambda$, so phantom universe will expand upto 
time $t_m$  given as 
$$ t_m = \frac{1}{A\sqrt{24\pi G}}\Big[t_0 +
2{(2\lambda)}^{(1/2) -\alpha}\sqrt{1-\frac{\rho_0}{2\lambda}}
2F_1\Big(\frac{1}{2},\frac{1}{2}+\alpha,\frac{3}{2},1-\frac{\rho_0}{2\lambda}\Big)\Big]
\eqno(3.5)$$
with $t_0$ being the present time.
Moreover, from (3.2), it is obtained that
$$ 3(1-\alpha)Aln(a_m/a_0)={(2\lambda)}^{1-\alpha}-\rho_0^{1-\alpha},
\eqno(3.6)$$  
where $ a_m = a(t_m).$ This equation shows that if $\alpha \gtrless
1, 2\lambda >\rho_0$ as $a_m >a_0.$

From (2.16) and (3.4), we obtain {\em effective phantom divide} at time
$$ t = t_{\rm phd} = \frac{1}{A\sqrt{24\pi G}}\Big[t_0 + 2{(2\lambda)}^{(1/2) -\alpha}\Big\{\sqrt{1-\frac{\rho_0}{2\lambda}} 2F_1\Big(\frac{1}{2},\frac{1}{2}+\alpha,\frac{3}{2},1-\frac{\rho_0}{2\lambda}\Big)$$
$$-\sqrt{1-\frac{\rho_{\rm phd}}{2\lambda}} 2F_1\Big(\frac{1}{2},\frac{1}{2}+\alpha,\frac{3}{2},1-\frac{\rho_{\rm phd}}{2\lambda}\Big)\Big\}\Big] , \eqno(3.7) $$ 
 
Inequalilties (2.15) and (2.17) show that for $\rho$ satisfying
$$  \rho <  \lambda \Big[2 +  3 A \rho^{(\alpha - 1)}
\Big\{1-\frac{\rho}{\lambda}\Big\}  \Big], \eqno(3.8)$$
SEC will be violated. It means that the universe will accelerate till $\rho$
will obey (3.8). But as $\rho$ will grow more and it wil satisfy
$$  \rho >  \lambda \Big[2 +  3 A \rho^{(\alpha - 1)}
 \Big\{1-\frac{\rho}{\lambda}\Big\}  \Big], \eqno(3.9)$$
SEC will not be violated. It means that the universe will decelerate when
 $\rho$ will  obey the inequality (3.9). It shows a transition from
 acceleration to deceleration at $\rho  = \rho_{\rm tr}$ given by the equation
$$  \rho_{\rm tr} =  \lambda \Big[2 +  3 A \rho_{\rm tr}^{(\alpha - 1)}
 \Big\{1-\frac{\rho_{\rm tr}}{\lambda}\Big\}  \Big]. \eqno(3.10)$$

Connecting (3.4) and (3.10), we obtain that this transition will take
place at time
$$ t = t_{\rm tr} = \frac{1}{A\sqrt{24\pi G}}\Big[t_0 +
2{(2\lambda)}^{(1/2) -\alpha}\Big\{\sqrt{1-\frac{\rho_0}{2\lambda}}
2F_1\Big(\frac{1}{2},\frac{1}{2}+\alpha,\frac{3}{2},1-\frac{\rho_0}{2\lambda}\Big)$$    
$$-\sqrt{1-\frac{\rho_{\rm tr}}{2\lambda}}
  2F_1\Big(\frac{1}{2},\frac{1}{2}+\alpha,\frac{3}{2},1-\frac{\rho_{\rm
      tr}}{2\lambda}\Big)\Big\}\Big] . \eqno(3.11)$$

If $\alpha = 1$, (3.3) integrates to
$$ \rho = \Big[\frac{1}{2\lambda} + \Big\{\sqrt{\frac{1}{\rho_0}
  -\frac{1}{2\lambda}} -A \sqrt{6 \pi G}(t - t_0)\Big\}^2\Big]^{-1},
\eqno(3.12)$$
 where $\rho_0< 2\lambda$ is the current energy density of the
DE.  It shows that phantom energy density will
increase with time, which is consistent with results of GR- based theory. 

From (2.2) and (3.12) it is obtained that
$$ H^2 = \frac{8\pi G}{3}\frac{\Big\{\sqrt{\frac{1}{\rho_0} -\frac{1}{2\lambda}} -A \sqrt{6 \pi G}(t - t_0)\Big\}^2}{\Big[\frac{1}{2\lambda} + \Big\{\sqrt{\frac{1}{\rho_0} -\frac{1}{2\lambda}} -A \sqrt{6 \pi G}(t - t_0)\Big\}^2\Big]^2}   \eqno(3.13) $$ 
The solution of (3.13) is 
$$ a(t) = a_0\rho_0^{-\frac{1}{3}}\Big[\frac{1}{2\lambda} + \Big\{\sqrt{\frac{1}{\rho_0} -\frac{1}{2\lambda}} -A \sqrt{6 \pi G}(t - t_0)\Big\}^2\Big]^{-\frac{1}{3}}       \eqno(3.14) $$
Using  $\rho=2\lambda$, in (3.12), it is obtained that phantom era will end at time $$ t_e = t_0 + \frac{1}{A \sqrt{6 \pi G}}\sqrt{\frac{1}{\rho_0} -\frac{1}{2\lambda}}  \eqno(3.15)$$
In this case, (2.4) reduces to
$$\frac{\ddot a}{a}=\frac{4\pi
  G\rho}{3}\Big[3A\Big(1-\frac{\rho}{\lambda}\Big) +  2
\Big(1-\frac{\rho}{2\lambda}\Big)\Big] ,   \eqno(3.16) $$ 
which yields 
$$\ddot a\gtreqless 0 \hspace{.5in} when \hspace{.5in} \frac{\rho}{\lambda} \lesseqgtr \frac{3 A+2}{1+3A}  \eqno(3.17)$$
(3.12) and (3.16) show that $\ddot a = 0$ when 
$$\frac{3A+1}{(3A+2)\lambda}=\frac{1}{2\lambda} + \Big\{\sqrt{\frac{1}{\rho_0} -\frac{1}{2\lambda}} -A \sqrt{6 \pi G}({t_v}_m - t_0)\Big\}^2 .\eqno(3.18)$$
It yields time for transition from acceleration to deceleration as
$${t_v}_m=t_0 + \frac{1}{A\sqrt{6\pi G}}\Big[\sqrt{\frac{1}{\rho_0}-\frac{1}{2\lambda}} -\sqrt{\frac{3A}{2(3A+2)\lambda}}\Big] \eqno(3.19)$$
From eq.(3.12), it is also obtained that $\ddot a < 0$ when $2\lambda \geqslant \rho >{(3A+2)\lambda}/{3A+1}$. It means that during the time interval
$$t_0 + \frac{1}{A\sqrt{6\pi
    G}}\sqrt{\frac{1}{\rho_0}-\frac{1}{2\lambda}}\geqslant t>t_0 +
\frac{1}{A\sqrt{6\pi G}}\Big[\sqrt{\frac{1}{\rho_0}-\frac{1}{2\lambda}}
-\sqrt{\frac{3A}{2(3A+2)\lambda}}\Big] \eqno(3.20a) $$ 
phantom universe will decelerate in case $\alpha=1$. It gives deceleration period $$ \sqrt{\frac{M_p^2}{4(3A+2)\pi G\lambda}}, \eqno(3.20b)$$
which depends on magnitude of negative brane tension. There is no deceleration
    if $\lambda \gg M_p^2$. So role of brane tension is very crucial here\cite{sks07}

\smallskip

\[
  \underline{Case II: f(\rho) =
  \frac{{A\rho^\alpha}}{{\sqrt{1-\frac{\rho}{2\lambda}}}}} 
\] 
In this case, connecting (2.3) and (2.5),we obtain
$$\dot\rho - 3A\frac{\dot a}{a}\frac{\rho^\alpha}{\sqrt{1-\rho/2\lambda}} = 0,
\eqno(3.21)$$ 
which integrates to
$$ 3A ln\Big(\frac{a}{a_0}\Big)=\frac{1}{3}
2^{2-\alpha}\lambda^{1-\alpha}\Big[{\Big(1-\frac{\rho_0}{2\lambda}\Big)}^{3/2}2F_1\Big(\frac{3}{2},\alpha,\frac{5}{2},1-\frac{\rho_0}{2\lambda}\Big)$$     
$$-{\Big(1-\frac{\rho}{2\lambda}\Big)}^{3/2}2F_1\Big(\frac{3}{2},\alpha,\frac{5}{2},1-\frac{\rho}{2\lambda}\Big)\Big]  ,   
 \eqno(3.22) $$ 
 where $\rho_0\leqslant\rho\leqslant 2\lambda $.

(2.2), (2.3) and (2.5) yield                                                                     
$$\dot\rho - 3A\sqrt{\frac{8\pi G}{3}}\rho^{\alpha +\frac{1}{2}} = 0
\eqno(3.23) $$ 
Exact solution of this equation is obtained as
$$  \rho = \Big[\rho_0^{(1-2\alpha)/2} + \frac{1}{2}(1- 2\alpha)A \sqrt{6\pi
  G}(t-t_0)\Big]^{{2}/{(1 - 2\alpha)}}.  \eqno(3.24)$$
Thus, (3.22) and (3.24) yield the scale factor $a(t)$ as function of time $t$.

As in RS-II model expansion stops at $\rho = 2 \lambda$, here phantom driven universe will expand upto time $t_m$ given as
$$t_m = t_0 + \frac{2}{A(1-2\alpha)\sqrt{6\pi G}}\Big[(2\lambda)^{(1-2\alpha)/2}- \rho_0^{(1-2\alpha)/2} \Big]   \eqno(3.25) $$

In this case, WEC will be violated till
$$t < t_0 + \frac{2}{A(1-2\alpha)\sqrt{6\pi G}}\Big[\rho^{(1-2\alpha)/2}-
\rho_0^{(1-2\alpha)/2} \Big]   \eqno(3.26a) $$ 
and $\rho$ will satisfy the inequality (2.20). It is found that  WEC will not
be violated when
$$t > t_0 + \frac{2}{A(1-2\alpha)\sqrt{6\pi G}}\Big[\rho^{(1-2\alpha)/2}- \rho_0^{(1-2\alpha)/2} \Big] ,  \eqno(3.26b) $$
and $\rho$ will satisfy the inequality (2.21). So, {\em effective phantom
  divide} is obtained at time
$$t_{\rm phd} = t_0 + \frac{2}{A(1-2\alpha)\sqrt{6\pi G}}\Big[\rho_{\rm phd}^{(1-2\alpha)/2}- \rho_0^{(1-2\alpha)/2} \Big] ,  \eqno(3.26c) $$
where $\rho_{\rm phd}$ given by (2.24). 

The inequality (2.23) and (3.24) yield the time period during which SEC will be
violated. It shows that phantom fluid will behave effectively as quintessence
fluid due to brane-corrections in this case. So, universe will accelerate till
$$t < t_{\rm ae} = t_0 + \frac{2}{A(1-2\alpha)\sqrt{6\pi
    G}}\Big[\rho_{\rm ae}^{(1-2\alpha)/2}- \rho_0^{(1-2\alpha)/2} \Big] ,  \eqno(3.27a)
$$ 
where  
$$ \rho = \rho_{\rm ae} = \lambda \Big[2 +  \frac{3 A \rho^{(\alpha- 1/2)}}
  {\sqrt{1 - {\rho_{\rm ae}}/{2\lambda}} } \Big\{1-\frac{\rho_{\rm
  ae}}{\lambda}\Big\} 
  \Big]. \eqno(3.27b)$$ 

 Similarly, inequalities (2.25) and (3.27b) as well as (3.24) and (3.27a)
 yield the time period $t_{\rm ae} < t < t_m$, during which brane-corrections of RS-II
 model will be so effective that  neither SEC nor WEC will be  violated for phantom
 fluid. As a consequence, universe will decelerate during this period.

Like the case I, here too, we take $\alpha = 1$ as an example. In what
follows, above results are analyzed if $\alpha
= 1$. 
Using (3.26c) time for {\em effective phantom divide} will be obtained at
$$ t_{\rm phd} = t_0 + \frac{2}{A\sqrt{6\pi G}}\Big[\rho_0^{(-1/2}- \rho_{\rm phd}^{(1-2\alpha)/2} \Big] ,  \eqno(3.28a) $$
where 
$$ \rho_{\rm phd} = \frac{3 \lambda A}{1 + 3 A} .\eqno(3.28b) $$

According to (3.25), universe will expand till
$$t_m = t_0 + \frac{2}{A\sqrt{6\pi G}}\Big[\rho_0^{-1/2} - (2\lambda)^{-1/2}
\Big] .  \eqno(3.29) $$

(3.22) and (3.24) show that at time
$$ t_{\rm br} = t_0 + 2/A \sqrt{6 \pi G \rho_0}  \eqno(3.30)$$
 $\rho$ is divergent and $a(t)$ is complex. It is an unphysical situation.

Connecting (3.29) and (3.30), it is obtained that
$$ t_m =  t_{\rm br} - \frac{1}{A\sqrt{3\pi G \lambda}}. \eqno(3.31) $$
It shows that expansion of the uiverse will stop before encountering the
unphysical situation occurring at time $t_{\rm br}$ given by (3.30).

\smallskip

\[
  \underline{{\rm Case III} :  f(\rho) =
  {A\rho^{1/2}ln(\rho/\rho_0)}/{\sqrt{1-\frac{\rho}{2\lambda}}}} 
\]

In this case, using  (2.2),(2.3) and (2.5),we get 
$$\dot \rho -3A\sqrt{\frac{8\pi G}{3}}\rho ln\rho = 0,    \eqno(3.32) $$
 which integrates to
$$ ln(\rho/\rho_0)= \sqrt{24\pi G }(t-t_0) . \eqno(3.33)$$ 

The {\em effective phantom divide} is obtained at time
$$t = t_{\rm phd} = t_0 + \rho_{\rm phd}/ A \rho_0 \sqrt{24\pi G } .  \eqno(3.34)$$ 
It shows that, at $t < t_{\rm phd}$, the phantom fluid will violate WEC, but,at $t
> t_{\rm phd}$ WEC will not be violated. 

(2.30) shows that even SEC will be violated till $\rho < \rho_{\rm ae}$ and it
will not be violated when $\rho > \rho_{\rm ae}$, where
$$  \rho_{\rm ae} = \lambda \Big[2 + \frac{3 A \rho_{\rm ae}^{- 1/2}ln(\rho_{\rm ae}/\rho_0)}
  {\sqrt{1- {\rho_{\rm ae}}/{2\lambda}} } \Big\{1-\frac{\rho_{\rm ae}}{\lambda}\Big\} \Big] .  \eqno(3.35)$$

Phantom energy will acquire the value $\rho_{\rm ae}$ at time
$$ t_{\rm ae} = t_0 +  \rho_{\rm phd}/ A \rho_0 \sqrt{24\pi G } 
\eqno(3.36)$$  
being obtained from (3.33) and (3.35).

Like above cases, in this case, $t_m$ is obtained as
$$ t_{\rm m} = t_0 +  2 \lambda/ A \rho_0 \sqrt{24\pi G } 
\eqno(3.37)$$ 

Results (3.36) and (3.37) show that universe will accelerate till $ t_0 \le t <
t_{\rm ae}$ and will decelerate for $t_{\rm ae} < t < t_{\rm m}.$

\bigskip

\centerline{\bf 4. Summary}

In this paper, we analyze the behaviour of phantom fluid in RS-II model of
brane-gravity having negative brane-tension $\lambda$. Here, three cases of non-linear
equations of state for the phantom fluid are taken. It is found that, contrary
to RS-I model, in RS-II model, brane-corrections
make drastic changes in the behaviour of phantom fluid, which is
characterized by violation of WEC and accelerating universe ending up in
big-rip 
singularity in most of the models. Interestingly, RS-II model based phantom
cosmology is found different from the usual picture of phantom universe. Energy
conservation for phantom fluid yields that phantom energy density increases
as universe expands. Above results suggest that behaviour of phantom fluid will
change in the future universe as energy density will grow with
expansion. Here, the model of the future universe begins at time $t_0$ being the
present age of the universe and it stops expanding when phantom energy density
$\rho$ grows to $2\lambda$ by the time $t_{\rm m}$. The  above analysis shows
that, during the period $t_0 \le t < t_{\rm m}$, two transitions will take
place. The first one will take place at  $t_{\rm phd}$ being the time of transition from {\em violation
  of WEC} to {\em non-violation of WEC} and {\em violation of
  SEC}. The  second one will take place at $t_{\rm ae}$ being the time of
transition from {\em violation of SEC} to {\em non-violation of SEC}. These
transitions are caused by brane-corrections due to negative brane-tension in
RS-II model-based universe.  As a consequence,
it is found that 
the present model of the universe will accelerate during the period $t_0 \le t
< t_{\rm ae}$  and decelerate during the period $ t_{\rm ae} < t < t_{\rm
  m}$. Moreover, the model is free from big-rip problem. Thus, it is found
that the role of brane-tension is very crucial. When it is negative, it causes
drastic changes in the behaviour of phantom dark energy, but phantom fluid
has usual behaviour when brane-tension is positive.

\end{document}